\documentclass{ptapap}

\author{Viktor Khalack}[UoM]
\author{Francis LeBlanc}[UoM]
\author{Oleksandr Kobzar}[UoM]
\affil[UoM]{D\'epartement de physique et d'astronomie, Universit\'e de Moncton, Moncton, N.B., Canada E1A 3E9}

\title{VeSElkA: Vertical Stratification of Element Abundances in CP stars}

\begin{document}

\maketitle

\begin{abstract}

A portion of the upper main-sequence stars, called chemically peculiar (CP) stars, show significant abundance anomalies mainly due to atomic diffusion of chemical elements within the stellar atmospheres of these stars. Slowly rotating CP stars may have hydrodynamically stable atmospheres where the competition between the gravitational and radiative forces leads to atomic diffusion that can be responsible for the abundance peculiarities observed in CP stars. Recently, project VeSElkA (Vertical Stratification of Element Abundances) was initiated with the aim to detect and study the vertical stratification of element abundances in the atmospheres of CP stars. Some results from abundance analysis of several slowly rotating (V$\sin{i}$< 40 km/s) CP stars observed recently with ESPaDOnS are presented here. Signatures of vertical abundance stratification for several chemical elements have been found in the stellar atmospheres of HD~22920, HD~41076, HD~95608, HD~116235, HD~148330 and HD~157087.

\end{abstract}

\section{Introduction}

More than 10 percent of the upper main sequence stars show abnormally strong or weak line profiles detected in their spectra which can be explained in terms of enhanced or depleted abundances of several chemical elements in their stellar atmospheres (e.g. \citealt{Gomez+98}). These stars show peculiarities of chemical abundances with respect to their solar value and therefore were named chemically peculiar (CP) stars \citep{Preston+74}. \citet{Preston+74} has divided the known CP stars in four distinct classes: CP1 -- metallic-line stars (Am), CP2 -- stars of spectral class A with a significant magnetic field (Ap), CP3 -- mercury-manganese stars (HgMn), and CP4 -- helium weak stars. Later, the F-type stars with strong metallic lines (Fm) were added to the class CP1, and the chemically peculiar magnetic B-type stars (Bp) were added to the class CP2. The peculiar abundances of chemical elements are only related to the stellar atmospheres or to certain layers of the stellar interior (in AmFm stars) and does not reflect the chemical composition of the entire star.
A catalogue of known and presumed CP stars with detailed information on their global stellar parameters was published by \citet{Renson+Manfroid09}. After the launch and successful work of the Transiting Exoplanet Survey Satellite (\textit{TESS}), more extensive data for CP stars can be found in the \textit{TESS} Input Catalogue (TIC, \citealt{Stassun+19}). A catalogue of known magnetic CP stars was also compiled by \citet{Bychkov+03}. A detailed review and statistical analysis of observed abundance peculiarities in known CP stars were recently published by \citet{Ghazaryan+19}.

Slowly rotating and magnetic CP stars can possess hydrodynamically stable stellar atmospheres where atomic diffusion can lead to accumulation or depletion of chemical abundances at certain optical depths \citep{Michaud70}. In a hydrodynamically stable atmosphere atomic diffusion can be responsible for horizontal (existence of under- or overabundance patches) and vertical stratification of abundances in the CP stars (e.g. \citealt{Michaud19}). In magnetic CP (hereafter mCP) stars, the magnetic field of dipolar structure can significantly amplify atomic diffusion and cause the existence of overabundance patches in certain areas of stellar atmosphere \citep{Alecian+Stift10}. \citet{Romanyuk+14} have shown that the observed configuration of magnetic field and variability of line profiles do not change over several decades in the analyzed mCP stars. This observational fact argues in favour of the hypothesis that the stellar atmospheres of mCP stars are hydrodynamically stable.


From the analysis of spectral line profiles of $\beta$~CrB, \citet{Ryabchikova+03} have shown that abundances of Fe and Cr increase towards the deeper atmospheric layers. It appears that the light and the iron-peak elements are concentrated in the lower atmospheric layers of some mCP stars \citep{Ryabchikova+04}. Meanwhile, the rare-earth elements (for example, Pr and Nd) are usually pushed into the upper atmosphere of Ap stars \citep{Mashonkina+05,Khalack+17}. Stratification of abundances with optical depth has been found by \citet{Castelli+17} for HD~6000 from the analysis of its ultraviolet spectra. While these studies have proven the existence of vertical abundance stratification in hydrodynamically stable atmospheres of CP stars, they do not provide enough data to carry out a statistical analysis to find out
why only a part of the upper main sequence stars shows a chemical peculiarity.

\section{Project VeSElkA}

To search for the signatures and to study the abundance stratification of chemical species with optical depth in the atmospheres of CP stars, we have initiated project VeSElkA \citep{Khalack+LeBlanc15a,Khalack+LeBlanc15b}, which stands for Vertical Stratification of Element Abundances. Slowly rotating (V$\sin{i}$< 40 km/s) CP stars of the upper main sequence are selected for our study using the catalogue of Ap, HgMn and Am stars of \citet{Renson+Manfroid09}. This limitation for the rotational velocity is imposed with the aim to increase the probability for a star to have a hydrodynamically stable atmosphere, where the atomic diffusion mechanism can result in vertical stratification of element abundances. A low value of V$\sin{i}$ also leads to narrow and mostly unblended line profiles in the observed spectra, which is beneficial for our abundance analysis. Rotational periods for some magnetic CP stars were derived from the analysis of magnetic field measurements, 
and from the analysis of flux variations \citep{Kobzar+19,David-Uraz+19,Sikora+19} measured by \textit{TESS} \citep{Ricker+15} (see Section~\ref{tess}).

\begin{figure}[t]
    \includegraphics[angle= -90,width=\textwidth]{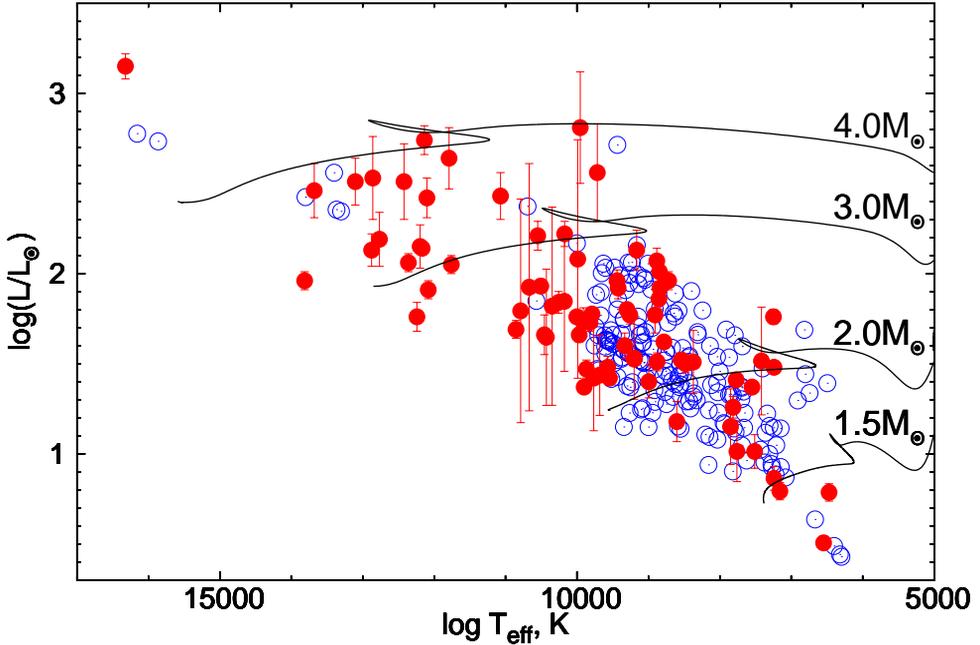}
    \caption{The HR diagram for the presumably singular slowly rotation CP stars observed in the frame of Project VeSElkA  (filed circles) and those detected with \textit{TESS} telescope and planned for observation with ESPaDOnS, NARVAL and HERMES (open circles). The continuous lines show evolutionary tracks calculated for stars of 1.5, 2.0, 3.0 and 4.0 $M_{\odot}$ assuming Z=0.014 and Y=0.273 \citep{Chen+14}.}
    \label{fig_HR}
\end{figure}

\begin{figure}[t]
    \includegraphics[angle= -90,width=\textwidth]{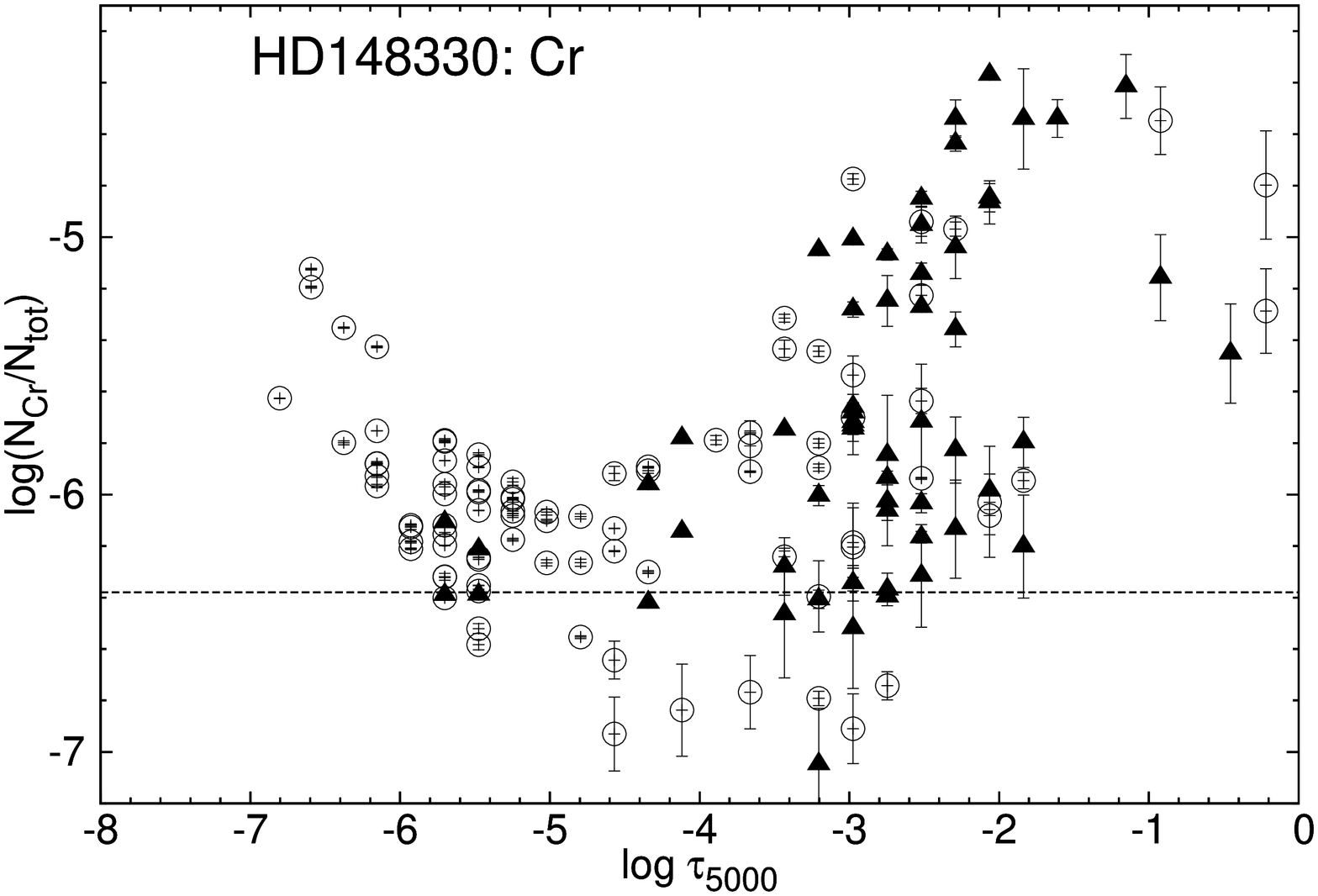}
    \caption{The variation of chromium abundance with optical depth in HD148330 \citep{Khalack+17}. Each symbol represents the data obtained from the analysis of a distinct line profile that belongs to the singly ionized chromium (open circles) or to its neutral atoms (filled triangles). The dashed line represents the solar chromium abundance \citep{Grevesse+10}.}
    \label{fig_strat}
\end{figure}


The high-resolution and high signal-to-noise spectra of more than 80 slowly rotating CP stars were obtained\footnote{Based on observations obtained at the Canada-France-Hawaii Telescope (CFHT) which is operated by the National Research Council of Canada, the Institut National des Sciences de l'Univers of the Centre National de la Recherche Scientifique of France, and the University of Hawaii, at the Telescope Bernard Lyot (TBL) at Observatoire du Pic du Midi, CNRS/INSU and Université de Toulouse, and at the Mercator Telescope, operated on the island of La Palma by the Flemmish Community, at the Spanish Observatorio del Roque de los Muchachos of the Instituto de Astrof\'{i}sica de Canarias. 
} during the last decade with the spectropolarimetres ESPaDOnS (CFHT) and NARVAL (TBL), and the spectrometer HERMES (telescope Mercator).
To estimate the effective temperature, gravity and metallicity of stars, new grids of stellar atmosphere models and corresponding fluxes have been calculated \citep{Husser+13,Khalack+LeBlanc15a} employing the code PHOENIX \citep{Hauschildt+97}. For the majority of CP stars observed in the frame of the VeSElkA project, we have determined their effective temperature, surface gravity and metallicity by fitting the Balmer line profiles with the help of the code FITSB2 \citep{Napiwotzki+04} and compared them with the estimates of the stellar global parameters derived from photometry \citep{Hauck+Mermilliod98,Netopil+08}.
On the HR diagram, the position of CP stars studied in the frame of the VeSElkA project is indicated by filled circles (see Fig.~\ref{fig_HR}).

Knowing the $T_{\rm eff}$, $\log{g}$ and metallicity we have calculated models of stellar atmospheres with the help of version 15 of the code PHOENIX \citep{Hauschildt+97} for certain CP stars assuming a homogeneous distribution of element abundances in their atmospheres. A stellar atmosphere model is required to carry out abundance analysis using the ZEEMAN2 radiative transfer code \citep{Landstreet88}. The code ZEEMAN2 was modified to automatically  analyze a sequence of several hundred line profiles \citep{Khalack+Wade06,Khalack+07}.
In this study, we presume that the core of the line profile is formed mainly at line optical depth $\tau_{line}$=1, which corresponds to a particular layer of the stellar atmosphere. Taking into account that the analyzed lines usually have different lower energy levels and oscillator strengths, their cores are generally formed at different optical depths $\tau_{5000}$. Employing this methodology, we can therefore track the vertical distribution of an element's abundance from the analysis of a large number (more than ten) of line profiles that belong to one or two ions of this element \citep{Khalack+08}. Spectra of a dozen CP stars have already been analyzed to search for vertical stratification of element abundances with optical depth in their stellar atmospheres.

We have found that in HD~22920 only silicon and chromium have a tendency to increase their abundance towards the deeper atmospheric layers \citep{Khalack+Poitras15}. Evidence of vertical stratification of iron and chromium abundances were found in the stellar atmospheres of HD~95608 and HD~116235 \citep{Khalack+14,LeBlanc+15}, and in HD~41076 and HD~148330 \citep{Khalack+17} (see Fig.~\ref{fig_strat}). \citet{Ndiaye+18} have found vertical stratification of phosphorus abundance in HgMn stars HD~53929 and HD~63975, while \citet{Khalack18} has reported detection of a significant increase of C, S, Ca, Ti, Sc, V, Cr, Mn, Fe, Co, Ni and Zr abundances towards the deeper atmospheric layers of HD~157087. \citet{Kashko+19} have found that Na, Al, Si, Fe, Zn, and Sr appear to be significantly overabundant in the stellar atmosphere of HD~63401. However, not all studied stars show signatures of vertical abundance stratification of chemical species \citep{LeBlanc+15}.


\section{Analysis of \textit{TESS} data}
\label{tess}

The space telescope \textit{TESS} was launched by NASA in April 2018 with the purpose of detecting exoplanets \citep{Ricker+15}, but can be also used to study internal stellar photometric variabilities. In 2019, \textit{TESS} finished the survey of the southern hemisphere (13 sectors) and provided short-cadence observations (with the 2~min accumulation time per measurement) for a subset of more than 200,000 stars. Each sector was observed during approximately 27 days and covers around 20,000 stars selected for the short-cadence observations. Some of these stars are present in multiple sectors and their time series were observed during a much longer period than 27 days.


Based on the analysis of \textit{TESS} data\footnote{Photometric data presented in this paper were obtained from the Mikulski Archive for Space Telescopes (MAST). STScI is operated by the Association of Universities for Research in Astronomy, Inc., under NASA contract NAS5-2655.} in the first four sectors, \citet{David-Uraz+19} and \citet{Sikora+19} have recently published the lists of rotationally variable CP stars. Rotational photometric variability is usually observed in CP stars with overabundance patches present in their
stellar atmospheres, and can be explained in terms of the oblique magnetic rotator (OMR) model \citep{Stibbs50}. In this case, we assume that a high-amplitude periodic variability in the light curve with the lowest frequency and the associated first harmonics correspond to the stellar rotation from which the rotational period may be derived.

We employed a high-performance autonomous computing procedure named TESS-AP (recently developed by V. Khalack) to carry out periodic analysis of short-cadence light curves obtained by \textit{TESS}, detect stellar pulsations and select stars with the desired type of photometric variability. This procedure carries out autonomous reduction of time series, performs periodic analysis, extracts additional data for studied targets from the public astronomical databases (TIC, SIMBAD, and GALAH), and stores the derived information in a final report \citep{Khalack+19}.
Taking into account that atomic diffusion is effective in a hydrodynamically stable stellar atmospheres 
we were searching for late B, early and late A and F type slowly rotating (P > 2d) CP/mCP stars with overabundance patches that can cause detectable (by \textit{TESS}) photometric variability assuming the OMR model. Based on the estimates of rotational periods of CP stars observed by \textit{TESS} in the sectors 1-13, a list of over a thousand of slowly rotating CP/mCP stars suitable for spectroscopic follow-up with the spectropolarimeters ESPaDOnS, NARVAL and spectrometer HERMES was selected. Using the global stellar parameters available for these targets in the TIC database \citep{Stassun+19} we plotted their position on the HR diagram (see open circles in Fig.~\ref{fig_HR}). The list also contains several early B-type hot stars. From the Fig.~\ref{fig_HR}, one can see that for some of those targets we have already obtained high-resolution and high signal-to-noise spectra and carried out an abundance analysis (see, for example \citealt{Kashko+19}).
\citet{Kobzar+19} have selected 8 relatively bright mCP stars from this list to study the rotational variability of their light curves and of their mean longitudinal magnetic field measurements.

TESS-AP is also capable of compiling a list of stars which show $\delta$~Scuti or roAp type pulsations. Simulation of stellar pulsations (relying on the large frequency separation) detected in HD~27463, which simultaneously shows properties of $\delta$~Scuti and roAp stars, allowed to correctly estimate its global stellar parameters using the spectroscopic constraints obtained from the analysis of Balmer line profiles \citep{Khalack+19}. It was also shown that the spectroscopic estimates of effective temperature and metallicity are required to break degeneracies amongst fundamental parameters when performing forward seismic modelling \citep{Aerts+18}.
Seismic modelling of the detected pulsation modes and large frequency separation in the studied CP stars is planned to be carried out using the codes MESA \citep{Paxton+19} and GYRE \citep{Townsend+Teitler13}. MESA is designed to simulate a grid of stellar structure and evolution models, while GYRE is used to calculate linear adiabatic pulsation frequencies for each model. The frequencies of detected pulsation modes will be fitted with the theoretical ones to find the best fit of global stellar parameters (see \citealt{Khalack+19}).


\section{Discussion}

More than 80 presumably singular slowly rotation CP stars have already been observed in the frame of the VeSElkA project. A sample of approximately a thousand slowly rotating CP/mCP candidates is compiled using the \textit{TESS} observations of the southern hemisphere. Employing the TESS-AP procedure we plan to increase this sample by adding the candidates observed by \textit{TESS} in the northern hemisphere as well.
A number of relatively bright CP/mCP stars from this sample are selected for the spectropolarimeteric observations with ESPaDOnS (CFHT) and NARVAL (TBL), and for spectroscopic follow-up with instrument HERMES (Mercator telescope).

Spectral analysis of Balmer line profiles and of the line profiles of metals visible in the high-resolution spectra of studied stars will result in measurements of their effective temperature, surface gravity, metallicity, V$\sin{i}$, and radial velocity. Knowing the rotational period derived from the analysis of \textit{TESS} data and V$\sin{i}$, one may estimate an inclination angle, $i$, of the stellar rotation axis to the line of sight and the rotational velocity at the stellar equator, $V_e$. Statistical analysis of correlation of the derived values for $V_e$, $i$ with the estimates of magnetic field intensity on the magnetic poles of mCP stars may provide additional information on evolution of magnetic fields in main sequence stars (see \citealt{Landstreet+09}).

The values of global stellar parameters derived from the analysis of Balmer line profiles will be compared with their estimates obtained from the asteroseismology modelling using the codes MESA \citep{Paxton+19} and GYRE \citep{Townsend+Teitler13}. The TESS-AP procedure and asteroseismology modelling will also help to validate the stellar classification of the observed objects, as they appear in existing catalogues of variable stars \citep{Khalack+19}.

The use of a semiautomatic pipeline \citep{Khalack+17} for spectral analysis of CP/mCP stars observed in the frame of the project VeSElkA will result in a significant amount of new data on average abundance for the studied stars and will provide more data on vertical stratification of element abundances. The found patterns of abundance variations with optical depth are going to be used as input parameters for the PHOENIX code to calculate an empirical “self-consistent” model of the stellar atmosphere that considers vertical abundance stratification of studied elements.
These results will allow us to establish a correlation between the vertical abundance stratification and the effective temperature, the structure of magnetic field (when present), and the stellar age. These results will also contribute to improving the codes designed to calculate stellar atmosphere models that take into account vertical stratification of chemical elements in non-magnetic \citep{LeBlanc+09} and magnetic \citep{Stift+Alecian16} CP stars. The relative effect of other physical processes such as mass-loss and convection on atomic diffusion may also be better constrained.

\section*{Acknowledgments}

V.K. and F.L. acknowledge the support from the Natural Sciences and Engineering Research Council of Canada (NSERC).
O.K. and V.K. are thankful to the Facult\'{e} des \'{E}tudes Sup\'{e}rieures et de la Recherche and to the Facult\'{e} des Sciences de l'Universit\'{e} de
Moncton for financial support of this research. Some calculations were carried out on the supercomputer \textit{b\'{e}luga} of the \'{E}cole de technologie sup\'{e}rieure in Montreal, under the guidance of Calcul Qu\'{e}bec and Calcul Canada.
This paper includes data collected by the \textit{TESS} mission. Funding for the \textit{TESS} mission is provided by the NASA Explorer Program. This research used the SIMBAD database, operated at CDS, Strasbourg, France.

\bibliographystyle{ptapap}
\bibliography{Khalack_London}

\end{document}